\documentclass[preprint,aps,amsmath,superscriptaddress,nofootinbib]{revtex4}
\usepackage{bm}
\usepackage{epsfig}

\newcommand{\nn}{\nonumber} 

\newcommand{\bea}{\begin{eqnarray}}
\newcommand{\eea}{\end{eqnarray}}

\newcommand{\Hb}{\bar{H}}

\DeclareMathOperator{\tr}{tr}
\newcommand{\lrbd}{\overset{\leftrightarrow}{\partial}}
\newcommand{\ket}[1]{| #1 \rangle}
\newcommand{\bra}[1]{\langle #1 |}

\begin{document}



\title{Heavy Quark Symmetry Predictions for Weakly Bound B-Meson Molecules}

\author{Thomas Mehen\footnote{Electronic address: mehen@phy.duke.edu}}
\affiliation{Department of Physics, 
	Duke University, Durham,  
	NC 27708\vspace{0.2cm}}

\author{Joshua W. Powell\footnote{Electronic address:    jwp14@phy.duke.edu}}
\affiliation{Department of Physics, 
	Duke University, Durham,  
	NC 27708\vspace{0.2cm}}

\date{\today\\ \vspace{1cm} }


\begin{abstract}

Recently the Belle collaboration discovered two resonances, $Z_b(10610)$ and $Z_b(10650)$, that lie very close to the $B \bar{B}^*$ and $B^* \bar{B}^*$ thresholds, respectively. It is natural to suppose that these are molecular states of bottom and anti-bottom mesons. Under this assumption, we introduce an effective field theory for the $Z_b(10610)$ and $Z_b(10650)$, as well as similar unobserved states that are expected on the basis of heavy quark spin symmetry.  The molecules  are assumed to arise from short-range interactions that respect heavy quark spin symmetry. We use the theory to calculate line shapes in the vicinity of $B^{(*)} \bar{B}^{(*)}$ thresholds as well as two-body decay rates of the new bottom meson bound states. We derive new heavy quark spin symmetry predictions for the parameters appearing in the line shapes as well as the total and partial widths of the states.  

\end{abstract}

\maketitle

\newpage

Recently the BELLE collaboration observed two resonances, $Z_b(10610)$ and $Z_b(10650)$, in the decays $\Upsilon(5S) \to \Upsilon(nS) \pi^+ \pi^-$ ($n=1,2,$ or 3) and $\Upsilon(5S) \to h_b(mP) \pi^+ \pi^-$ ($m = 1$ or 2)~\cite{Collaboration:2011gja}. The $Z_b(10610)$ and $Z_b(10650)$ (which we will refer to as $Z_b$ and $Z_b^\prime$ below) have widths of about 15 MeV, and their masses lie a few MeV above the $B \bar{B}^*$ and $B^* \bar{B}^*$ thresholds, respectively. However, an analysis in Ref.~\cite{Cleven:2011gp} concludes that  an interpretation of the states as bound states lying below the $B^{(\ast)} \bar{B}^\ast$ threshold is still  consistent with the available data on the decays $\Upsilon(5S) \to Z_b^{(\prime)} \pi \to h_b \pi^+ \pi^-$. This conclusion depends on using line shapes for the $Z^{(\prime)}_b$ that account for the coupling of the $Z^{(\prime)}_b$ to the nearby $B^{(\ast)}\bar{B}^\ast$ thresholds rather than the Breit-Wigner form that was used in the experimental analysis.
The experimental analysis favors the quantum numbers $I^G(J^P) = 1^+ (1^+)$   for the $Z_b$ and $Z_b^\prime$ states. Arguments based on heavy quark spin symmetry \cite{Bondar:2011ev,Voloshin:2011qa} indicate that there should be similar states called $W_{b0}$ and $W^\prime_{b0}$ with quantum numbers $I^G(J^P)= 1^-(0^+)$, as well as possibly $W_{b1}$ and $W_{b2}$ with quantum numbers $1^-(1^+)$ and $1^-(2^+)$, respectively. 

In this paper, we will assume that these states are weakly bound molecules of heavy mesons. This is the approach adopted in Refs.~\cite{Yang:2011rp, Chen:2011zv, Zhang:2011jja}.  For alternative interpretations of these states as tetraquarks, see Refs.~\cite{Guo:2011gu, Ali:2011vy, Cui:2011fj}.  Ref.~\cite{Nieves:2011vw} uses existence of the X(3872) and arguments based on heavy quark symmetry to argue that molecular states in the bottom sector must exist.  Assuming the states are weakly bound molecules means they can be studied using a low energy effective field theory (EFT) that consists of nonrelativistic kinetic terms for the mesons and contact interactions whose coefficients are tuned to produce the bound states with energies close to threshold. A theory of this kind called XEFT has been developed for $X(3872)$, which is thought to be a shallow $S$-wave bound state of 
$D^0 \bar{D}^{*0} + \bar{D}^0 D^{\ast0}$ \cite{Fleming:2007rp, Fleming:2008yn, Braaten:2010mg, Mehen:2011ds}.  The purpose of this paper is to construct the analogous theory for the isovector $Z_b$, $Z_b^\prime$, and $W^{(\prime)}_{bJ}$ states. This theory is similar in structure to the pionless effective theory  used for very low energy nuclear physics \cite{Chen:1999tn,vanKolck:1998bw}. It will be used to derive line shapes for the resonances that are valid near the relevant $B^{(*)}\bar{B}^{(*)}$ thresholds as well as calculate the two-body decays of the resonances. The predicted line shapes and decay rates incorporate the constraints imposed by heavy quark symmetry. New predictions for the parameters in the line shapes and the total and partial widths of these states are obtained. Experimental tests of these predictions should aid in interpreting the newly discovered $Z_b$ and $Z_b^\prime$ states and searching for their partners.

The EFT of this paper can be applied when the relative momentum of the $B^{(*)} \bar{B}^{(*)}$ mesons is much smaller than the pion mass.  In two-nucleon scattering, a box diagram with two-pions is suppressed relative to one-pion exchange by $Q/\Lambda_{NN}$  where $Q \sim p \sim m_\pi$ and $1/\Lambda_{NN} =g_A^2 M_N/(8 \pi f_\pi^2) \approx 1/(300 \, {\rm MeV})$, where $M_N$ is the nucleon mass, $g_A$ is the nucleon axial coupling, and $f_\pi$ is the pion decay constant~\cite{Kaplan:1998tg}. Perturbative treatment of pions fails rather badly in two-nucleon systems \cite{Fleming:1999ee} when $p \geq m_\pi$. For  pion exchanges between a $B^*$ and $B^{(*)}$ meson, the expansion parameter is $Q/\Lambda_{BB}$, where $1/\Lambda_{BB} =g^2 M_B/(8 \pi f_\pi^2)$. Here 
$M_B$ is the $B$ meson mass and the axial coupling of heavy mesons, $g$, is between 0.5 and 0.7, which yields 160 MeV $\leq \Lambda_{BB} \leq$ 320 MeV. Since $\Lambda_{BB} \approx \Lambda_{NN}$ we expect that perturbative treatment of pions will fail in the B meson sector for $p\sim m_\pi$, but a pionless effective theory should work for $p\ll m_\pi$.

The Lagrangian we will use to describe low energy $B^{(*)} \bar{B}^{(*)}$ scattering is
\bea\label{Lag}
{\cal L} &=& {\rm Tr}[H^\dagger_a \left(i \partial_0 +\frac{\vec\nabla^2}{2 M}\right)_{ba} H_b] + \frac{\Delta}{4}{\rm Tr}[H^\dagger_a \,  \sigma^i \, H_a \, \sigma^i] \,  \label{lagrangian}\\
&+&{\rm Tr}[\Hb^\dagger_a \left(i \partial_0 +\frac{\vec\nabla^2}{2M}\right)_{ab}  \Hb_b] + \frac{\Delta}{4}{\rm Tr}[\Hb^\dagger_a \,  \sigma^i \, \Hb_a \, \sigma^i] \, \nn \\
&-&\frac{C_{00}}{4} \, {\rm Tr}[\bar{H}_a^\dagger  H_a^\dagger H_b \bar{H}_b ] 
- \frac{C_{01}}{4}  {\rm Tr}[\bar{H}^\dagger_a  \sigma^i H^\dagger_a H_b \sigma^i \bar{H}_b ]  \nn\\
&-&  \frac{C_{10}}{4}{\rm Tr}[\bar{H}^\dagger_a \tau^A_{aa^\prime} H^\dagger_{a^\prime} H_b \tau_{bb^\prime}^A \bar{H}_{b^\prime} ] 
- \frac{C_{11}}{4}  {\rm Tr}[\bar{H}^\dagger_a \tau^A_{aa^\prime}  \sigma^i H^\dagger_{a^\prime} H_b \tau_{bb^\prime}^A  \sigma^i \bar{H}_{b^\prime} ]   \nn \, .
\eea
Here $a$ and $b$ are $SU(2)$ isospin indices, isospin matrices are normalized as $\tau^A_{ab} \tau^B_{ba} =\delta^{AB}$, traces are over spin  indices which are not explicit, and $H_a (\bar{H}_a)$ is the heavy meson (heavy anti-meson) superfield. In terms of components, $H_a = P_a+ V^i \sigma^i$ and $\bar{H}_a =\bar{P}_a - \bar{V}^i_a  \sigma^i$, where 
$P_a \,(\bar{P}_a)$ and $V^i_a \, (\bar{V}^i_a)$ are the pseudoscalar and vector $\bar{B}$ ($B$) mesons, respectively.
The  transformation properties of the fields under heavy quark spin and other symmetries are given in Ref.~\cite{Fleming:2008yn}.
In Eq.~\eqref{Lag}, the mass $M$ in the kinetic terms is the spin-averaged $B$ meson mass, $M= (3 M_{B^*}+M_B)/4 = 5314$ MeV. 
The first three terms are the leading heavy hadron chiral perturbation theory Lagrangian of Refs. \cite{Wise:1992hn, Burdman:1992gh, Yan:1992gz},
written in the two-component notation of Ref.~\cite{Hu:2005gf}. The next three terms are the Lagrangian for anti-heavy mesons. 
The kinetic terms must be promoted to leading order to prevent pinch singularities in $B^{(*)} \bar{B}^{(*)}$ scattering, as is conventional in nonrelativistic theory. The heavy mesons and anti-heavy mesons interact via the remaining terms in the Lagrangian which are contact interactions that mediate $S$-wave heavy meson scattering. Contact interactions of this type first were written down in Ref.~\cite{AlFiky:2005jd}, where the operators considered are proportional 
to ${\rm Tr}[H^\dagger_a H_a] \, {\rm Tr}[\bar{H}_b^\dagger \bar{H}_b]$ and ${\rm Tr}[H^\dagger_a H_a \sigma^i] \,{\rm Tr}[\bar{H}^\dagger_b \sigma^i \bar{H}_b]$. It is easy to see that these operators can be written in terms of the single trace operators given above using Fierz transformations. In the case of $B^{(*)} \bar{B}^{(*)}$ scattering, one can classify states in terms of the total spin of the heavy quark and antiquark ($S_{Q\bar{Q}}$), and the total angular momentum ($S_{q\bar{q}}$) and isospin ($I$) of the light degrees of freedom. Since the light degrees of freedom in the $B$ and $\bar{B}$ meson are isodoublets and have spin-$1/2$, the possible states of the light degrees of freedom are: i)  $I = S_{q\bar{q}} = 0$, ii)  $I=0$ and $S_{q\bar{q}} = 1$, iii) $I=1$ and $S_{q\bar{q}} = 0$, or iv) $I = S_{q\bar{q}} = 1$. The four operators in Eq.~(\ref{Lag}) mediate $S$-wave scattering in each of these channels and the notation for the coefficients is that the operator with coefficient $C_{IS}$ mediates scattering in the isospin $I$ and spin $S=S_{q\bar{q}}$ channel. In the heavy quark limit scattering should be independent of $S_{Q\bar{Q}}$.

Ref.~{\cite{Bondar:2011ev} classified possible bound states of $B$ and $\bar{B}$ mesons  and concluded that there should be at least four and maybe six such isotriplet states.  The wavefunctions of these states in terms of their $S_{Q\bar{Q}}$ and $S_{q\bar{q}}$ quantum numbers are derived in Ref.~\cite{Voloshin:2011qa}.  (See also Eq.~\eqref{molecularcomponents} below.)   $W_{b1}$ and $W_{b2}$ are pure states with $S_{Q\bar{Q}} = S_{q\bar{q}} = 1$. The remaining states are mixtures of $S_{q\bar{q}} = 0$ or $1$ and $S_{Q\bar{Q}} = 0$ or 1. If the mechanism that leads to shallow bound states operates in the $S_{q\bar{q}} = 1$ channel or both $S_{q\bar{q}} = 0$ and $S_{q\bar{q}} = 1$ channels, one expects to find all six shallow bound states. If the mechanism operates only in the $S_{q\bar{q}} = 0$ channel, then only four shallow bound  states ($Z_b, Z_b^\prime,W_{b0}$, and $W_{b0}^\prime$) are expected. We will primarily focus on the former case, then comment on the latter at the end of the paper.

The interpolating fields for these states are given by (we will drop the subscript $b$ in what follows):
\bea\label{interpolating}
Z^{A \,i} &=&  \frac{1}{\sqrt{2}}(V^i_a \tau^A_{ab} \bar{P}_b - P_a \tau^A_{ab} \bar{V}^i_b) \\
Z^{\prime \,A \,i} &=&  \frac{i}{\sqrt{2}}\,\epsilon^{ijk} V^j_a \tau^A_{ab} \bar{V}^k_b  \nn \\
W^A_0 &=&   P_a \tau^A_{ab} \bar{P}_b \nn \\
W^{\prime A}_0 &=& \frac{1}{\sqrt{3}} V^i_a \tau^A_{ab} \bar{V}^i_b \nn \\
W_1^{A\,i} &=& \frac{1}{\sqrt{2}}  (V^i_a \tau^A_{ab} \bar{P}_b + P_a \tau^A_{ab} \bar{V}^i_b) \nn \\
W_2^{A\,\lambda} &=&  \epsilon^\lambda_{ij} V^i_a \tau^A_{ab} \bar{V}^j_b \, ,\nn
\eea
where $\epsilon^\lambda_{ij}$ is  a basis for symmetric traceless polarization vectors normalized as $\epsilon^\lambda_{ij} \epsilon^{\lambda^\prime}_{ij} = \delta^{\lambda \lambda^\prime}$, and $a$ and $b$  label flavor antifundamental and fundamental indices, respectively.  It is also possible to define isoscalar interpolating fields that are obtained from those in Eq.~(\ref{interpolating}) by dropping the index $A$ and replacing $\tau^A_{ab} \to \delta_{ab}/\sqrt{2}$. We will focus on isovector states in what follows, the generalization to isoscalars is straightforward. It is enlightening to rewrite the contact interactions in terms of these interpolating fields. For the isovector fields  these are 
\bea\label{contact}
{\cal L}_{\rm contact} &=& -2 \,C_{11} \left(W_1^{A\,i \,\dagger } W_1^{A\,i} + \sum_\lambda W_{2 \, \lambda}^{A \,\dagger} W^A_{2\, \lambda} \right) \\
&& -\frac{1}{2} \left(\begin{array}{cc} W^{A\,\prime \, \dagger}_0 & W_0^{A\,\dagger} \end{array}\right)
\left(\begin{array}{cc} 3 \,C_{10} +C_{11} & \sqrt{3}(C_{11} -C_{10}) \\
 \sqrt{3}(C_{11} -C_{10}) & C_{10} + 3 \,C_{11} \end{array}\right)
\left(\begin{array}{c} W^{A\,\prime}_0 \\  W^A_0  \end{array}\right) \nn \\
&&
- \left(\begin{array}{cc} Z^{\prime \,A\, i \, \dagger} & Z^{A\,i\,\dagger} \end{array}\right)
\left(\begin{array}{cc}  C_{10} +C_{11} &  C_{11} -C_{10} \\
C_{11} -C_{10} & C_{10} + C_{11} \end{array}\right)
\left(\begin{array}{c} Z^{\prime\,A \, i} \\  Z^{A\,i} \end{array}\right) \nn \\
&=&  -2 \,C_{11} \left(W^{A\, \dagger}_{0+} W^A_{0+} + Z^{A\, i\,\dagger}_+ Z^{A\, i}_+  +
W_1^{A\,i \,\dagger } W_1^{A\,i} + \sum_\lambda W_{2 \, \lambda}^{A \,\dagger} W^A_{2\, \lambda} \right) \label{diagonalized}\\
&& -2 \,C_{10} \left( W^{A\, \dagger}_{0-} W^A_{0-} + Z^{A\, i\,\dagger}_- Z^{A\, i}_- \right) \, ,\nn
\eea
where in the last line the interactions are  diagonalized  by defining the fields $W^A_{0+} = \frac{1}{2} W_0^{\prime \,A} + \frac{\sqrt{3}}{2}W^A_0$, $W^A_{0-} = \frac{\sqrt{3}}{2} W_0^{\prime \,A} - \frac{1}{2}W^A_0$ and $Z^A_{\pm} = \frac{1}{\sqrt{2}} (Z^A \pm Z^{A\,\prime})$.  The Lagrangian for isoscalar terms is obtained by dropping the superscripts $A$ and replacing $C_{1i} \to C_{0i}$, $i=0$ or $1$.  Though we have diagonalized the interactions in Eq.~\eqref{diagonalized}, we will not work in this basis because the $Z^A$ and $Z^{\prime \, A}$ are split by the hyperfine splitting, $\Delta = 46$ MeV, and the $W_0^A$ and $W^{\prime \, A}_0$ are split by $2 \Delta = 92$ MeV.

It is straightforward to calculate the T-matrix for $B^{(*)}\bar{B}^{(*)}$ scattering in these channels. For the $W^A_2$ channel we find 
\bea
T_{W_2}=  \frac{1}{-1/(2C_{11})- \Sigma_{B^*\bar{B}^*}(E)} \, ,
\eea
where $\Sigma_{B^*\bar{B}^*}(E)$ is computed from a one loop diagram containing nonrelativistic $B^*$ and $\bar{B}^\ast$ propagators of total energy $E$ and is given by:
\bea\label{sigma}
\Sigma_{B^*\bar{B}^*}(E) = \frac{M}{4\pi} \left(\Lambda -\sqrt{ M (2\Delta-E)-i \epsilon}\right) \, .
\eea
The energy $E$ is measured with respect to the $B \bar{B}$ threshold.
The linear divergence in Eq.~(\ref{sigma}) is cancelled by the coupling constant
\bea\label{C11}
C_{11} =  C_{11}(\Lambda) = \frac{2\pi}{M}\frac{1}{-\Lambda +\gamma_{11}} \, ,
\eea
so the T-matrix is given by
\bea
T_{W_2}=  \frac{4\pi}{M}\frac{1}{-\gamma_{11}+ \sqrt{ M (2\Delta - E)-i \epsilon}} \, .
\eea
The $T$-matrix has a bound state pole at $E=2\Delta -\gamma^2_{11}/M$ for $\gamma_{11} >0$. 
A similar calculation for the $W^A_1$   channel yields
\bea
T_{W_1}=  \frac{4\pi}{M}\frac{1}{-\gamma_{11}+ \sqrt{ M (\Delta - E)-i \epsilon}} \, ,
\eea
which has a bound state pole at $E = \Delta - \gamma^2_{11}/M$. If shallow bound states $W_{1}^A$ and $W_2^A$ exist, heavy quark symmetry predicts their binding energies to be the same. On the other hand, if $\gamma_{11} <0$, then there are no shallow bound states. In this case, heavy quark symmetry predicts the $S$-wave scattering length for $B$ meson scattering in these channels to be the same. 

For $Z^A$ and $Z^{\prime\, A}$ states we must solve a coupled channel problem. The $T$-matrix is given by 
\bea
T_Z^{-1} = - C_Z^{-1} - \Sigma_Z(E) \, ,
\eea
where $C_Z$ and $\Sigma_Z(E)$ are matrices given by
\bea
C_Z &=& \left(\begin{array}{cc}  C_{10} +C_{11} &  C_{11} -C_{10} \\
C_{11} -C_{10} & C_{10} + C_{11} \end{array}\right) \, , \\
\Sigma_Z(E)  &=& \left(\begin{array}{cc}  \Sigma_{B^*\bar{B}^*}(E) &  0  \\
0 &  \Sigma_{B \bar{B}^*}(E)  \end{array}\right)\\
&=&  \frac{M}{4\pi}
\left(\begin{array}{cc}  \Lambda-\sqrt{M(2\Delta-E)-i\epsilon} &  0 \\
0 &  \Lambda-\sqrt{M(\Delta-E)-i\epsilon} \end{array}\right) \, , \nn\\
\Sigma_{B \bar{B}^*}(E) &=& \frac{M}{4\pi} \left(\Lambda -\sqrt{ M (\Delta-E)-i \epsilon}\right) \, .
\eea 
The cutoff dependence in the $T$-matrix can be completely cancelled if the coupling $C_{10}$ has the same form as $C_{11}$ in Eq.~(\ref{C11}), i.e., if $C_{10}= C_{10}(\Lambda) = 2\pi/(M(-\Lambda +\gamma_{10}))$. For the $T$-matrix  we find 
\bea
T_Z &=& \left(\begin{array}{cc}  T_{Z^\prime Z^\prime} &  T_{Z^\prime Z}  \\
T_{Z Z^\prime}  &  T_{Z Z}  \end{array}\right) \, ,
\eea
with the components given by
\bea
T_{Z^\prime Z^\prime} &=&\frac{4 \pi}{M} \frac{-\gamma_+ +\sqrt{M(\Delta-E)-i\epsilon}}{(\gamma_+ -\sqrt{M(\Delta-E)-i\epsilon})(\gamma_+ -\sqrt{M(2\Delta-E)-i\epsilon}) -\gamma_-^2 } \\
T_{Z^\prime Z} &=& T_{Z Z^\prime} =\frac{4 \pi}{M} \frac{\gamma_- }{(\gamma_+ -\sqrt{M(\Delta-E)-i\epsilon})(\gamma_+ -\sqrt{M(2\Delta-E)-i\epsilon}) -\gamma_-^2 }
\\
T_{ZZ} &=&\frac{4 \pi}{M}
\frac{-\gamma_+ +\sqrt{M(2\Delta-E)-i\epsilon}}{(\gamma_+ -\sqrt{M(\Delta-E)-i\epsilon})(\gamma_+ -\sqrt{M(2\Delta-E)-i\epsilon}) -\gamma_-^2 } \, ,
\eea
where $\gamma_\pm = (\gamma_{11}\pm \gamma_{10})/2$. 

Solving the analogous problem in the $W_0^\prime$ and $W_0$ channels, we obtain
\bea
T_{W^\prime W^\prime} &=&\frac{4 \pi}{M} \frac{-\gamma^{W}_+ +\sqrt{-ME-i\epsilon}}
{(\gamma^W_+ -\sqrt{-ME-i\epsilon})(\gamma^{W^\prime}_+ -\sqrt{M(2\Delta-E)-i\epsilon}) -(\gamma^W_-)^2 } \\
T_{W^\prime W} &=& T_{W W^\prime} = \frac{4 \pi}{M} \frac{\gamma^W_- }
{(\gamma^W_+ -\sqrt{-ME-i\epsilon})(\gamma^{W^\prime}_+ -\sqrt{M(2\Delta-E)-i\epsilon})-(\gamma^W_-)^2 }
\\
T_{W W} &=&\frac{4 \pi}{M}
\frac{-\gamma^{W^\prime}_+ +\sqrt{M(2\Delta-E)-i\epsilon}}
{(\gamma^W_+ -\sqrt{-ME-i\epsilon})(\gamma^{W^\prime}_+ -\sqrt{M(2\Delta-E)-i\epsilon}) -(\gamma^W_-)^2 } \, ,
\eea
where $\gamma^W_+ = (\gamma_{10} +3\gamma_{11})/4$,  $\gamma^{W^\prime}_+ = (3\gamma_{10} + \gamma_{11})/4$,
and $\gamma^W_- = \sqrt{3}(\gamma_{11}-\gamma_{10})/4 = \sqrt{3}\gamma_-/2$. 

These amplitudes are valid near the relevant threshold.  Specifically, if we take the molecular state to have a binding energy of $\gamma^2/M$, then for $\gamma \ll m_\pi$ we can expand the amplitudes in powers of $\gamma/\sqrt{M\Delta}$, where $\sqrt{M\Delta} = 494\,{\rm MeV}$.  For example, the $Z_b^\prime$ state in the vicinity of the $B^\ast \bar{B}^\ast$ threshold will have energy $E = 2\Delta - \gamma^2/M$.  Expanding the denominator in the expression for $T_{Z^\prime Z^\prime}$ to leading order in $\gamma$, we find
\bea
T_{Z^\prime Z^\prime} = \frac{4 \pi}{M} \frac{1}{-\gamma_+ + \gamma +O(\gamma^2/\sqrt{\Delta M})} \, ,
\eea
The pole in the amplitude is at $\gamma \simeq \gamma_+$ or $E\simeq 2\Delta - \gamma^2_+/M$, corresponding to a $Z^\prime$ mass $m_{Z^\prime} = 2 M_{B^*} -\gamma^2_+/M$.  For the other states the binding energies are $\gamma_A^2/M$, where A=$Z,Z^\prime,W_0,W_0^\prime,W_1$ or $W_2$, and are given by
\bea\label{gammarelations}
\gamma_Z &=& \gamma_{Z^\prime} = \gamma_+ \\
\gamma_{W_1} &=& \gamma_{W_2} = \gamma_{11} \nn\\
\gamma_{W_0} &=& \frac{\gamma_{10} + 3\gamma_{11}}{4} = \frac{\gamma_Z + \gamma_{W_1}}{2} \nn \\
\gamma_{W_0^\prime} &=& \frac{3\gamma_{10} +  \gamma_{11}}{4} = \frac{3\gamma_Z - \gamma_{W_1}}{2} \nn \, .
\eea
These relations between the binding momenta are consequences of heavy quark symmetry.

We can incorporate the effects of decays of these resonances on the line shapes by explicitly violating unitarity.  If the decays to other states also respect heavy quark spin symmetry, then we expect that incorporating these decays will just give imaginary components in the couplings of Eq.~\eqref{Lag}.  So we promote $C_{00}$, $C_{01}$, $C_{10}$ and $C_{11}$ to complex values, which also means $\gamma_{00}$, $\gamma_{01}$, $\gamma_{10}$ and $\gamma_{11}$ are complex.  For each of them, we can write $\gamma_{IS} = -1/a_{IS} + i\Gamma_{IS}/2$, where $a_{IS}$ is the scattering length and $\Gamma_{IS}$ is the total width of the bound state in the $IS$ channel.  The relations in Eq.~\eqref{gammarelations} will still hold since they are a consequence of heavy quark spin symmetry and now give relationships among their imaginary components, i.e., the total widths.  One prediction
\bea
\Gamma_{11} = \Gamma[W_1]=\Gamma[W_2]=\frac{3}{2}\Gamma[W_0] -\frac{1}{2}\Gamma[W_0^\prime] \, ,
\label{voloshinsrelation}
\eea
was first derived in Ref.~\cite{Voloshin:2011qa}.  In addition we also find that
\bea
\Gamma_+ = \frac{1}{2}(\Gamma_{11} + \Gamma_{10}) = \Gamma[Z]=\Gamma[Z^\prime]=\frac{1}{2}(\Gamma[W_0] +\Gamma[W_0^\prime] ) \, .\label{decayraterels}
\eea
The relation $\Gamma[Z] =\Gamma[Z^\prime]$ was first derived in Ref.~\cite{Bondar:2011ev}, while the last equality of Eq.~(\ref{decayraterels}) is new.
One could derive this result, as well as similar predictions for partial decay rates, from the following decomposition of the wavefunction of the molecular states in terms of their components of definite $S_{Q\bar{Q}} \otimes S_{q\bar{q}}\,$\cite{Voloshin:2011qa}:
\bea\label{wvfns}
W_2: && 1_{Q\bar{Q}} \otimes 1_{q\bar{q}}\Big|_{J = 2}\label{molecularcomponents} \\
W_1: && 1_{Q\bar{Q}} \otimes 1_{q\bar{q}}\Big|_{J = 1} \nn\\
W^\prime_{b0}: && \frac{\sqrt{3}}{2}\,0_{Q\bar{Q}} \otimes 0_{q\bar{q}} + \frac{1}{2}\,1_{Q\bar{Q}} \otimes 1_{q\bar{q}}\Big|_{J = 0} \nn\\
W_0: && \frac{\sqrt{3}}{2}\,1_{Q\bar{Q}} \otimes 1_{q\bar{q}}\Big|_{J = 0} - \frac{1}{2}\,0_{Q\bar{Q}} \otimes 0_{q\bar{q}}\nn\\
Z^\prime: && \frac{1}{\sqrt{2}}\,0_{Q\bar{Q}} \otimes 1_{q\bar{q}} - \frac{1}{\sqrt{2}}\,1_{Q\bar{Q}} \otimes 0_{q\bar{q}}\nn\\
Z: && \frac{1}{\sqrt{2}}\,0_{Q\bar{Q}} \otimes 1_{q\bar{q}} + \frac{1}{\sqrt{2}}\,1_{Q\bar{Q}} \otimes 0_{q\bar{q}}\nn \,.
\eea
The molecular states inherit their widths from those of their constituent states with definite $S_{Q\bar{Q}} \otimes S_{q\bar{q}}$.  Since the same constituent state appears in multiple molecules, by restricting to decays which are sensitive only to one choice of $S_{Q\bar{Q}} \otimes S_{q\bar{q}}$, one can arrive at relations among molecular decays.  For $S_{Q\bar{Q}} \otimes S_{q\bar{q}} = 1 \otimes 1$, the result is Eq.~\eqref{voloshinsrelation}.  To derive the prediction in Eq.~\eqref{decayraterels}, consider the two $S_{Q\bar{Q}}^z = 0$ spin configurations of the two heavy quarks:
\bea
\ket{S_{Q\bar{Q}} = 1, S_{Q\bar{Q}}^z = 0} = \frac{1}{\sqrt{2}}\,(\ket{\!\uparrow\downarrow} + \ket{\!\downarrow\uparrow}), \quad \ket{S_{Q\bar{Q}} = 0, S_{Q\bar{Q}}^z = 0} = \frac{1}{\sqrt{2}}\,(\ket{\!\uparrow\downarrow} - \ket{\!\downarrow\uparrow}).
\eea
If $\hat S^z_Q$ is the operator that measures the magnetic quantum number of the heavy quark only, then it is clear that
\bea
2\hat S_Q^z \ket{1,0} = \ket{0,0}\,, \quad 2\hat S_Q^z \ket{0,0} = \ket{1,0}\,.
\eea
Now let $M_{s, s^\prime}$ denote the interpolating fields with $S_{Q\bar{Q}} = s$ and $S_{q\bar{q}} = s^\prime$, with all other indices labeling other quantum numbers suppressed for compactness.  Then, it follows from the above that
\bea
[2\hat{S}_Q^z, M^\dagger_{1,0}] = M^\dagger_{0,0}, \quad [2\hat{S}_Q^z, M^\dagger_{0,0}] = M^\dagger_{1,0}.
\eea
In what follows, $h$ stands for any bottomonium state with allowed quantum numbers and $S_{Q\bar{Q}} = 0$, and $\ell$ is any allowed configuration of light hadrons.  We define $\ket{\widetilde{h}^i} = 2S_Q^i\ket{h}$, which means $\widetilde{h}$ is a bottomonium state with $S_{Q\bar Q} = 1$.
Then for the matrix element mediating the transition $W_0^\prime \to h\ell$, we find
\bea
 \bra{h \ell}W_0^{\prime\,\dagger}\ket{0} &=& \frac{\sqrt{3}}{2}\,\bra{h \ell}M^\dagger_{0,0}\ket{0}\label{HQSSrels}\\
&=& \frac{\sqrt{3}}{2}\,\bra{h \ell}[2\hat S_Q^z,M^\dagger_{1,0}]\ket{0}\nn\\
&=& \frac{\sqrt{3}}{2}\,\bra{h \ell}2\hat S_Q^z M^\dagger_{1,0}\ket{0}\nn\\
&=& \frac{\sqrt{3}}{2}\,\bra{\widetilde{h}^{(S^z=0)} \ell}M^{\dagger\, (S^z=0)}_{1,0}\ket{0}\nn\\
&=& \sqrt{\frac{3}{2}}\,\bra{\widetilde{h}^{(S^z=0)} \ell}Z^{\dagger\, (S^z=0)}\ket{0}\nn\\
&=& -\sqrt{\frac{3}{2}}\,\bra{\widetilde{h}^{(S^z=0)} \ell}Z^{\prime\,\dagger\,(S^z=0)}\ket{0}\nn
\eea
Rotational symmetry can be used to extend this result to other values of $S^z$ and moreover implies that
\bea
\bra{\widetilde{h}^{(S^z=m)} \ell}Z^{\dagger\,(S^z=m^\prime)}\ket{0} \propto \delta^{m m^\prime}.
\eea
Applying the result for $h = \eta_b$, which means $\widetilde{h} = \Upsilon$, it follows that
\bea
|{\cal M}(W_0^\prime \rightarrow \eta_b \ell)|^2 = \frac{3}{2}\times\frac{1}{3}\sum_{\rm spins} |{\cal M}(Z \rightarrow \Upsilon\ell)|^2.
\eea
A similar analysis can be performed starting with a $W_0$ instead.  We find
\bea
\Gamma[W_0 \rightarrow \eta_b \ell] \,:\, \Gamma[W_0^\prime \rightarrow \eta_b \ell] \,:\, \Gamma[Z \rightarrow \Upsilon \ell] \,:\, \Gamma[Z^\prime \rightarrow \Upsilon \ell] &=& \frac{1}{2} \,:\, \frac{3}{2} \,:\, 1 \,:\, 1\,.\label{decayrates1}
\eea
This result is valid in the extreme heavy quark limit, in which $\Upsilon$, $\eta_b$, $Z^{(\prime)}$ and $W_J^{(\prime)}$ are all degenerate.~\footnote{The degeneracy of $Z^{(\prime)}$ and $W^{(\prime)}$ is \emph{not} a consequence of heavy quark spin symmetry, e.g., the binding energies predicted from Eq.~(\ref{gammarelations}) are not the same.  However, these corrections to the masses of these resonances are $O(\gamma^2/M)$ and are expected to be a few MeV or less, and hence small compared to mass differences inherited from the hyperfine splittings of their constituent mesons. This lack of degeneracy is because the $Z^{(\prime)}$ and $W^{(\prime)}$ are linear combinations of members of different heavy quark spin multiplets, cf. Eq.~(\ref{wvfns}).  The multiplets of heavy quark spin symmetry are the $(Z^A_+, W^A_+)$ and $(Z^A_-,W^A_-)$ defined in the last line of Eq.~(\ref{contact}) and in terms of these states heavy quark spin symmetry predicts  $\Gamma[W^- \to \eta_b\ell] =\Gamma[Z^-\to \Upsilon\ell]$ and $\Gamma[W^+ \to \chi_b\ell] =\Gamma[Z^+\to h_b \ell]$. }  
  In reality decay rates will also depend on the available phase space, which can be sensitive to the hyperfine splittings in each of the multiplets.  For example, in the decays to single pions calculated below, the rates are multiplied by a prefactor of $E_\pi^2 \, k_\pi$ or $k_\pi^3$ and these kinematic prefactors introduce large corrections to Eq.~\eqref{decayrates1} when the splittings between multiplets are not significantly larger than the hyperfine splittings of the multiplets.  For decays to $P$-wave bottomonia, a similar derivation shows that
\bea
\Gamma[W_0 \rightarrow \chi_{b1} \ell] \,:\, \Gamma[W_0^\prime \rightarrow \chi_{b1} \ell] \,:\, \Gamma[Z \rightarrow h_b \ell] \,:\, \Gamma[Z^\prime \rightarrow h_b \ell] &=& \frac{3}{2} \,:\, \frac{1}{2} \,:\, 1\,:\, 1\,,\label{decayrates2}
\eea
in the heavy quark limit.  Eqs.~\eqref{decayrates1} and \eqref{decayrates2} together imply the relationships among total decay rates in Eq.~\eqref{decayraterels} assuming decays to quarkonia dominate the decays of the molecular states.

One can explicitly check the claimed relationships among decay rates by computing the rates to final states with one pion using HH$\chi$PT.  To obtain these rates, we will add to the Lagrangian of Eq.~\eqref{lagrangian} the following interactions
\bea \label{HHchiPT_lagrangian}
{\cal L}_{\rm HH\chi PT} &=& g \,{\rm Tr}[\Hb^\dagger_a \,\sigma^i \Hb_b] A^i_{ab}  - g\, {\rm Tr}[H^\dagger_a  H_b \, \sigma^i] A^i_{ba} \\
&+& \frac{1}{2} g_{\pi \Upsilon,n} \tr[\Upsilon_n^\dagger H_a \bar H_b]A^0_{ab} + \frac{1}{2} g_{\Upsilon,n}\tr[\Upsilon_n^\dagger H_a \sigma^j i\lrbd_j \bar H_a] + \text{h.c.}\nn\\
&+& \frac{1}{2}g_{\pi\chi,n}\tr[\chi^\dagger_{n,i} H_a \sigma^j \bar{H}_b]\epsilon_{ijk}A^k_{ab} + \frac{i}{2}\,g_{\chi,n} \tr[\chi^\dagger_{n,\,i} H_a \sigma^i \bar{H}_a] + \text{h.c.} \, .
\nn\eea
The first line above gives the $\pi B^{(\ast)}$ and $\pi\bar{B}^{(\ast)}$ interactions.  The next two lines are the interactions of bottomonium states with the B-mesons.  Heavy quark spin symmetry groups the $S$- and $P$-wave bottomonium states into the multiplets
\bea
\Upsilon_n &=& \sigma_i \Upsilon^i(nS) + \eta_b(nS),\\
\chi^i_n &=& \sigma_\ell\Big(\chi_{b2}^{i\ell}(nP) + \frac{1}{\sqrt{2}}\epsilon^{i\ell m}\chi_{b1}^m(nP) + \frac{1}{\sqrt{3}}\delta^{i\ell}\chi_{b0}(nP)\Big) + h^i_b(nP).\nn
\eea
Note that the Lagrangian is heavy quark spin symmetric except for the hyperfine splitting terms proportional to $\Delta$, so the symmetry is restored in the limit $\Delta = 0$.  From Eq.~\eqref{HHchiPT_lagrangian}, one can compute the decay rates using the methods given in Refs.~\cite{Fleming:2007rp, Fleming:2008yn, Braaten:2010mg, Mehen:2011ds}.  The decay rates to $S$-wave bottomonia and a single pion are listed below.  We will drop the quantum number $n$, labeling the radial excitation level, but it is important to keep in mind that the coupling constants $g_{\pi\Upsilon}$, $g_{\pi\chi}$, $g_\Upsilon$ and $g_\chi$ will be different for distinct multiplets. The decay rates to $S$-wave bottomonia and a single pion are
\bea
\Gamma[W_0 \rightarrow \pi \eta_b] &=& \frac{m_\eta k_\pi E_\pi^2}{8\pi m_{W_0} f_\pi^2} \Big[g_{\pi \Upsilon} - 2g  g_\Upsilon \,\frac{k_\pi^2}{E_\pi(E_\pi + \Delta)}\Big]^2\times{\cal O}_1\label{swavedecay}\\
\Gamma[W_0^\prime \rightarrow \pi \eta_b] &=& \frac{3 m_\eta k_\pi E_\pi^2}{8\pi m_{W_0^\prime} f_\pi^2}\Big[g_{\pi \Upsilon} - 2 g g_\Upsilon \frac{k_\pi^2}{E_\pi^2}\Big(1 + \frac{1}{3}\frac{\Delta}{E_\pi - \Delta}\Big)\Big]^2\times{\cal O}_2\nn\\
\Gamma[Z \rightarrow \pi \Upsilon] &=& \frac{m_\Upsilon k_\pi E_\pi^2}{4\pi m_Z f_\pi^2} \Big[\Big[g_{\pi \Upsilon} - 2 g g_\Upsilon \frac{k_\pi^2}{E_\pi^2}\Big(1 - \frac{\Delta}{3}\frac{E_\pi - 2\Delta}{E_\pi^2 - \Delta^2}\Big)\Big]^2 + \frac{2}{9}\Big[g g_\Upsilon \frac{k_\pi^2}{E_\pi^2}\frac{\Delta}{E_\pi - \Delta}\Big]^2\Big]\times{\cal O}_3\nn\\
\Gamma[Z^\prime \rightarrow \pi \Upsilon] &=& \frac{m_\Upsilon k_\pi E_\pi^2}{4\pi m_{Z^\prime} f_\pi^2} \Big[\Big[g_{\pi \Upsilon} - 2 g g_\Upsilon \frac{k_\pi^2}{E_\pi^2}\Big(1 + \frac{1}{3}\frac{\Delta}{E_\pi - \Delta}\Big)\Big]^2 + \frac{2}{9}\Big[g g_\Upsilon \frac{k_\pi^2}{E_\pi^2}\frac{\Delta}{E_\pi - \Delta}\Big]^2 \Big]\times{\cal O}_4\nn \, .
\eea
The decays to $P$-wave bottomonia and a single pion are
\bea
\Gamma[W_0 \rightarrow \pi \chi_{b1}] &=& \frac{m_{\chi_{b1}}\,k_\pi^3}{8\pi m_{W_0} f_\pi^2}\Big[g_{\pi\chi} + \frac{2 g g_\chi}{E_\pi + \Delta}\Big]^2\times {\cal O}_1\label{pwavedecay}\\
\Gamma[W_0^\prime \rightarrow \pi \chi_{b1}] &=& \frac{m_{\chi_{b1}}\,k_\pi^3}{24\pi m_{W^\prime_0} f_\pi^2}\Big[g_{\pi\chi} + \frac{2 g g_\chi}{E_\pi - \Delta}\Big]^2\times {\cal O}_2\nn \\
\Gamma[Z \rightarrow \pi h_b] &=& \frac{m_{h_b}\,k_\pi^3}{12\pi m_{Z} f_\pi^2}\Big[g_{\pi\chi} + g g_\chi\Big(\frac{1}{E_\pi} + \frac{1}{E_\pi + \Delta}\Big)\Big]^2\times {\cal O}_3\nn \\
\Gamma[Z^\prime \rightarrow \pi h_b] &=& \frac{m_{h_b}\,k_\pi^3}{12\pi m_{Z^\prime} f_\pi^2}\Big[g_{\pi\chi} + g g_\chi\Big(\frac{1}{E_\pi} + \frac{1}{E_\pi - \Delta}\Big)\Big]^2\times {\cal O}_4\nn\\
\Gamma[W_1 \rightarrow \pi \chi_{b0}] &=& \frac{m_{\chi_{b0}} k_\pi^3}{18\pi m_{W_1} f_\pi^2}\,\Big[g_{\pi\chi} + 2gg_\chi\Big(\frac{3}{4}\frac{1}{E_\pi - \Delta} + \frac{1}{4}\frac{1}{E_\pi + \Delta}\Big)\Big]^2\times{\cal O}_5\nn\\
\Gamma[W_1 \rightarrow \pi \chi_{b1}] &=& \frac{m_{\chi_{b1}} k_\pi^3}{24\pi m_{W_1} f_\pi^2}\,\Big[g_{\pi\chi} + \frac{2gg_\chi}{E_\pi}\Big]^2\times{\cal O}_5\nn\\
\Gamma[W_1 \rightarrow \pi \chi_{b2}] &=& \frac{5m_{\chi_{b2}} k_\pi^3}{72\pi m_{W_1} f_\pi^2}\,\Big[g_{\pi\chi} + \frac{2gg_\chi}{E_\pi + \Delta}\Big]^2\times{\cal O}_5\nn\\
\Gamma[W_2 \rightarrow \pi \chi_{b1}] &=& \frac{m_{\chi_{b1}} k_\pi^3}{24\pi m_{W_2} f_\pi^2}\,\Big[g_{\pi\chi} + \frac{2gg_\chi}{E_\pi - \Delta}\Big]^2\times{\cal O}_6\nn\\
\Gamma[W_2 \rightarrow \pi \chi_{b2}] &=& \frac{m_{\chi_{b2}} k_\pi^3}{8\pi m_{W_2} f_\pi^2}\,\Big[g_{\pi\chi} + \frac{2gg_\chi}{E_\pi}\Big]^2\times{\cal O}_6\nn
\, .
\eea
Decays not listed here can only proceed through higher-derivative interactions which are suppressed compared to those listed.  For compactness, we have defined the following non-perturbative factors:
\bea
{\cal O}_1 &=& \frac{1}{3}|\bra{0} P_a \tau^A_{ab} \bar{P}_b \ket{W_0^A}|^2\\
{\cal O}_2 &=& \frac{1}{3}|\bra{0} \tfrac{1}{\sqrt{3}} V^i_a\tau^A_{ab} \bar{V}^i_b \ket{W_0^{\prime \, A}}|^2\nn\\
{\cal O}_3 &=& \frac{1}{9}|\bra{0} \tfrac{1}{\sqrt{2}}(V^i_a \tau^A_{ab} \bar{P}_b - P_a \tau^A_{ab} \bar{V}^i_b) \ket{Z^{A\,i}} |^2\nn\\
{\cal O}_4 &=& \frac{1}{9}|\bra{0} \tfrac{1}{\sqrt{2}}\,i\epsilon^{ijk} V^j_a \tau^A_{ab} \bar{V}^k_b \ket{Z^{\prime\,A\,i}} |^2\nn\\
{\cal O}_5 &=& \frac{1}{9}|\bra{0} \tfrac{1}{\sqrt{2}} (V^i_a \tau^A_{ab} \bar{P}_b + P_a \tau^A_{ab} \bar{V}^i_b) \ket{W_1^{A\,i}}|^2\nn\\
{\cal O}_6 &=& \frac{1}{15}|\bra{0} \epsilon_\lambda^{ij} V^i_a \tau^A_{ab} \bar{V}^j_b \ket{W_2^\lambda}|^2\nn \,  .
\eea
Each of these has a prefactor of $1/3$ to prevent overcounting when summing over the isospin states $A$, and an additional prefactor of $1/(2J+1)$ for a molecular state with spin $J$ to prevent overcounting when summing over the spin polarization.  In the limit of exact heavy quark spin symmetry, using arguments like that given in Eq.~\eqref{HQSSrels} one can show that the ${\cal O}_i$'s are all equal.

The ratios of the decay rates in Eq.~\eqref{swavedecay} and Eq.~\eqref{pwavedecay} match the predicted results from Eq.~\eqref{decayrates1} and 
 Eq.~\eqref{decayrates2}, respectively, when $\Delta = 0$ and $E_\pi$ is the same for all decays, which will be the case when the heavy quark spin symmetry multiplets
$(\Upsilon_b,\eta_b)$, $(\chi_{bJ},h_b)$, and the $Z^{(\prime)}$ and $W_0^{(\prime)}$ are degenerate. Our explicit calculations of the decay rates allow us to incorporate important corrections to heavy quark spin symmetry predictions that come from phase space and kinematic factors. In HH$\chi$PT there are two mechanisms that contribute to the decay of the bound states. There is a short-distance process, mediated by the contact interactions  $g_{\pi \Upsilon}$ and $g_{\pi \chi}$, in which the $B^{(*)} \bar{B}^{(*)}$ transition to the final state quarkonium and pion  at a point.
If this process dominates, the predicted ratios of rates are the heavy quark symmetry predictions in Eqs.~(\ref{decayrates1}) and Eqs.~(\ref{decayrates2}) weighted by factors of $E_\pi^2 k_\pi$ and $k_\pi^3$, respectively.  There is also a long-distance process $B^{(*)} \bar{B}^{(*)} \to B^{(*)} \bar{B}^{(*)}\pi$ followed by coalesence of the $B$ and $\bar{B}$ meson into the  final state quarkonium through the couplings $g_\Upsilon$ and $g_\chi$. These processes lead to a more complicated dependence on the pion energy.  When these processes dominate, the dependence is well approximated by $k_\pi^5/E_\pi^2$ since in all cases $E_\pi \gg \Delta$. 

For decays to $S$-wave bottomonium, the two processes appear at the same order in the expansion. If contact interactions dominate, which is obtained when the dimensionless ratio of couplings,  $\lambda_\Upsilon = g g_\Upsilon/g_{\pi \Upsilon} \ll 1$, then the ratios of partial decay rates are predicted to be 
\bea\label{ratio1}
\Gamma[W_0 \rightarrow \pi\eta_b(3S)]\,&:&\,\Gamma[W^\prime_0 \rightarrow \pi\eta_b(3S)]\,:\,\Gamma[Z \rightarrow \pi\Upsilon(3S)]\,:\,\Gamma[Z^\prime \rightarrow \pi\Upsilon(3S)]\nonumber\\
&=& 0.26\,:\,2.0\,:\,0.62\,:\,1 \hspace*{2.3cm} (\lambda_\Upsilon  = 0) \, ,
\eea
where all partial decay rates have been normalized to the rate for $\Gamma[Z^\prime \rightarrow \pi\Upsilon(3S)]$. 
In the opposite limit, 
\bea\label{ratio2}
\Gamma[W_0 \rightarrow \pi\eta_b(3S)]\,&:&\,\Gamma[W^\prime_0 \rightarrow \pi\eta_b(3S)]\,:\,\Gamma[Z \rightarrow \pi\Upsilon(3S)]\,:\,\Gamma[Z^\prime \rightarrow \pi\Upsilon(3S)]\nonumber\\
&=& 0.12\,:\,2.1\,:\,0.41\,:\,1 \hspace*{2.3cm} (|\lambda_\Upsilon|  = \infty) \, .
\eea
\begin{figure}[t]
\centerline{
	\includegraphics[height=4cm]{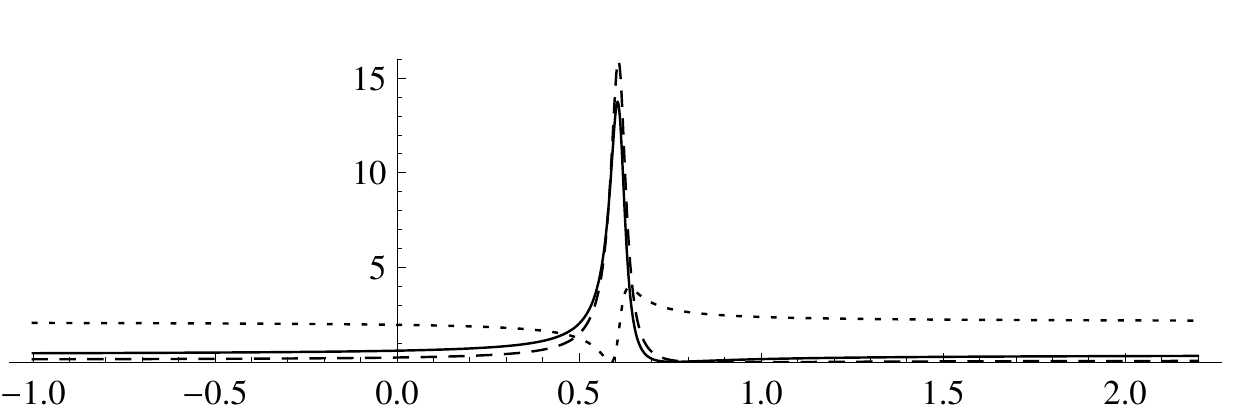}
}
\caption{Ratio of $\Gamma[W_0 \rightarrow \eta_b \pi]$ (dashed), $\Gamma[W^\prime_0 \rightarrow \eta_b \pi]$ (dotted) and $\Gamma[Z \rightarrow \Upsilon \pi]$ (solid) to \\ $\Gamma[Z^\prime \rightarrow \Upsilon \pi]$ as functions of $g\,g_\Upsilon / g_{\pi \Upsilon}$}
\label{ratioplots}
\end{figure}
\begin{figure}[t]
\centerline{
	\includegraphics[height=4cm]{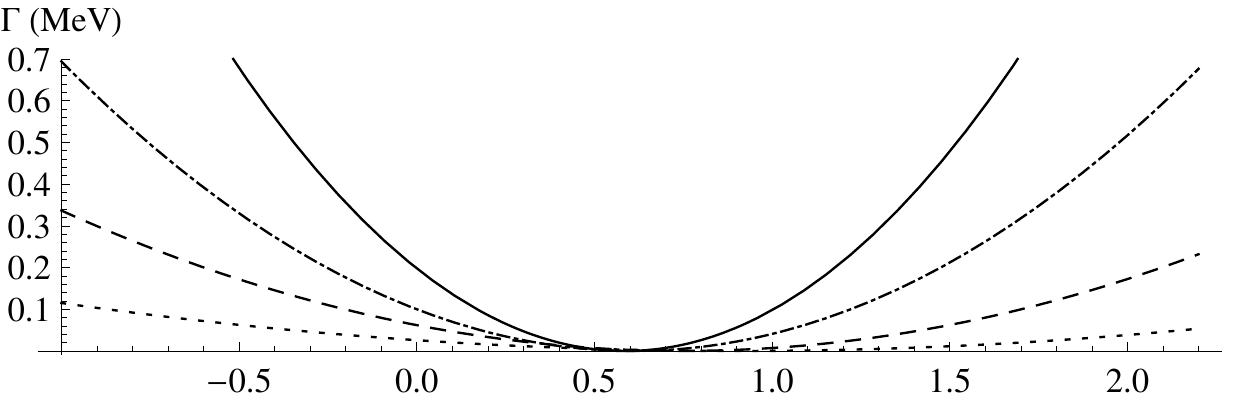}
}
\caption{Partial decay rates $\Gamma[W_0 \rightarrow \eta_b(3S)\pi]$ (dotted), $\Gamma[W^\prime_0 \rightarrow \eta_b(3S)\pi]$ (solid), $\Gamma[Z \rightarrow \Upsilon(3S)\pi]$ (dashed) and $\Gamma[Z^\prime \rightarrow \Upsilon(3S)\pi]$ (dashed-dotted) as functions of $g\,g_\Upsilon / g_{\pi\Upsilon}$ for one possible choice of currently undetermined parameters (${\cal O}\,g_{\pi\Upsilon}^2 = 10^{-3}$).}
\label{absplots}
\end{figure}
In computing these ratios, we have assumed ${\cal O}_1 = {\cal O}_2 = \cdots = {\cal O}_6$ holds without significant corrections from symmetry violating terms.  The values used for the masses of the bottomonium decay products, $m_{\Upsilon(3S)} = 10355\,{\rm MeV}$ and $m_{\eta_b(3S)} = 10328\,{\rm MeV}$, were determined in Ref.~\cite{Ebert:2002pp} using a relativistic quark model.

The results do not differ greatly because ratios of the kinematic factors $k_\pi^3$ and $k_\pi^5/E_\pi^2$ are not that different. For intermediate values of $\lambda_\Upsilon$ the results can depend rather dramatically on $\lambda_\Upsilon$ when $\lambda_\Upsilon \approx 0.6$, as seen in Fig.~\ref{ratioplots}. However, 
this wild variation occurs because of a cancellation between the two processes mentioned above. For these values of $\lambda_\Upsilon$, all four partial rates are highly suppressed, as shown in Fig.~\ref{absplots}. In this figure, we have chosen ${\cal O}_i = (100\, {\rm MeV})^3$ and $g_{\pi \Upsilon}^2 = 1.0 \, {\rm GeV}^{-3}$ and calculated the rates as a function of $\lambda_{\Upsilon}$. Our choices of  ${\cal O}_i$ and $g_{\pi \Upsilon}^2$ are based on naive dimensional analysis and are only intended to be estimates that should be accurate within a factor of 10. For this particular choice of ${\cal O}_i$ and $g_{\pi \Upsilon}^2$, the branching fraction for $Z\to \Upsilon \pi$ is less than 1\%, for any value of $\lambda_\Upsilon$ close enough to 0.6 that the ratios deviate significantly from those given in Eqs.~(\ref{ratio1}) and (\ref{ratio2}). Though this branching fraction has not been measured, the observation of this decay leads us to expect 
that the parameter $\lambda_\Upsilon$ does not take on values where such cancellations suppress the decay rates. Therefore, we  expect that 
experimental measurement of the ratios will yield results close to those in Eq.~(\ref{ratio1}) or Eq.~(\ref{ratio2}).

For decays to $P$-wave bottomonium, the processes mediated by contact interactions are suppressed in the power counting. 
The relative importance of leading order to contact interaction mediated processes is controlled by the dimensionful parameter
$g_{\pi \chi}/g_{\chi}$ which we expect to be $\approx 1 \, {\rm GeV}^{-1}$. In the limit $g_{\pi \chi}/g_{\chi} = 0$ we find
\bea\label{ratio3}
\Gamma[W_0 \rightarrow \pi \chi_{b1}(2P)]\,&:&\,\Gamma[W^\prime_0 \rightarrow \pi \chi_{b1}(2P)]\,:\,\Gamma[Z \rightarrow \pi h_b(2P)]\,:\,\Gamma[Z^\prime \rightarrow \pi h_b(2P)]\nonumber\\
&=& 0.72\,:\,0.57\,:\,0.66\,:\,1 \hspace*{1.7cm} (g_{\pi\chi}/g_\chi = 0\, {\rm GeV^{-1}}) \, ,
\eea 
and 
\bea\label{ratio4}
\Gamma[W_1 \rightarrow \pi \chi_{bJ}(2P)]\,&:&\,\Gamma[W_2 \rightarrow \pi \chi_{bJ}(2P)]\,:
 \frac{3}{2}\,\Gamma[W_0 \rightarrow \pi \chi_{b1}(2P)] - \frac{1}{2}\,\Gamma[W_0^\prime \rightarrow \pi \chi_{b1}(2P)]\nonumber\\
&=& 0.81\,:\,1\,:\,0.43 \hspace*{1.7cm} (g_{\pi\chi}/g_\chi = 0\, {\rm GeV^{-1}}) \, .
\eea
The masses used to compute these ratios are $m_{\chi_{b0}(2P)} = 10233\,{\rm MeV}$, $m_{\chi_{b1}(2P)} = 10255\,{\rm MeV}$, $m_{\chi_{b2}(2P)} = 10269\,{\rm MeV}$ and $m_{h_{b(2P)}} = 10261\,{\rm MeV}$.  The first three of these came from the Particle Data Group and the last is computed in Ref.~\cite{Ebert:2002pp}.  For $g_{\pi \chi}/g_{\chi} \approx 1\,{\rm GeV}^{-1}$ these ratios change by only a few percent.  For $g_{\pi \chi}/g_{\chi} \approx -(300\,{\rm GeV})^{-1}$, there is a cancellation between contact and leading order diagrams which suppresses the total rates and which leads to modifications of the ratios similar to what was discussed above  in the decays to $S$-wave bottomonium. Again the observation of $Z^\prime \to h_b(2P) \pi$ disfavors such a suppression of the decay rates.  Such a cancellation between leading order and next-to-leading order contributions is also inconsistent with the power counting of the theory. 

Throughout this paper, it is assumed that all six isovector molecular states exist and are loosely bound.  As stated earlier, it is also possible that binding occurs in the $S_{q\bar{q}} = 0$ channel only so that only the $W_0^{(\prime)}$ and $Z^{(\prime)}$ bound states exist.  In this case, the treatment of this paper is still applicable.  One simply takes the binding momentum $\gamma_{11} = 1/a_{11}$ to be small but negative.  In this case, resumming both the $C_{10}$ and $C_{11}$ interactions remains necessary.  There is also the possibility that in these channels there are no shallow bound states or large scattering lengths.  In this case, one can keep the summation in both channels as presented in this paper but tune $\gamma_{11}$ so there is no large scattering length.  It should also be possible to sum only the strong, binding interaction and give a perturbative treatment to the weaker interaction.  The $T$ matrices computed using the latter approach should correspond to a power series expansion of those in this paper.  We will save any investigation along these lines for future work.

In this paper we introduced the Lagrangian describing heavy quark spin symmetric $S$-wave contact interactions among observed and hypothesized isovector B meson molecules.  We derive the line shapes of these states in the vicinity of their respective $B^{(\ast)}\bar{B}^{(\ast)}$ thresholds, including coupled channel effects where mixture of states is possible.  By doing so, we have arrived at relationships among the binding energies and decay rates of the molecular states.  Some relationships among the widths of the $W^{(\prime)}_{bJ}$ and $Z^{(\prime)}$ states derived in this paper appear in Ref.~\cite{Voloshin:2011qa}, while the prediction in Eq.~\eqref{decayraterels} is new.  A confirmation of these predictions by explicit calculation of partial widths for strong two-body decays to $S$- and $P$-wave bottomonia using HH$\chi$PT was performed.  This allowed us to compute corrections to the earlier predictions which arise from differences in the kinematics between the various processes.  Tests of these predictions will aid in interpreting the new states.  Future work could include an extension of these results to the isoscalar sector of B meson molecules, a detailed look at radiative decays, as well as a determination of currently unknown parameters $\lambda_\Upsilon$ and $g_\chi / g_{\pi\chi}$ using the angular distribution of decay products.
\acknowledgments 

This work was supported in part by the  Director, Office of Science, Office of High Energy
Physics, of the U.S. Department of Energy under Contract No. DE-AC02-05CH11231U.S (SF), Office of Nuclear Physics, of the U.S. Department of Energy under grant numbers DE-FG02-06ER41449 (S.F.), 
DE-FG02-05ER41368 (T.M.), and DE-FG02-05ER41376 (T.M.).
 

\end{document}